\documentclass[twocolumn,aps,floats,floatfix]{revtex4}
\usepackage{graphicx,amssymb}

\begin{document}
\title{ Subwavelength Transportation of Light with Atomic Resonances}

\author{Siu-Tat Chui$^1$, Shengwang Du$^2$, and Gyu-Boong Jo$^{2}$}

\affiliation{$^1$Bartol Research Institute, University of Delaware, Newark, DE
19716, USA\\
$^2$ Department of Physics, The Hong Kong University of Science and Technology, Clear Water Bay, Kowloon, Hong Kong, China}

\begin{abstract}
We propose and investigate  a new type of optical waveguide made by an array of atoms without involving conventional Bragg scattering or total internal reflection. A finite chain of atoms collectively coupled through their intrinsic resonance supports a propagating mode with minimal radiative loss when the array spacing $a$ is around 0.6$\lambda_0/2\pi$ where $\lambda_0$ is the wavelength of the nearly resonant optical transition. We find that the transportation is robust with respect to position fluctuation and remains possible when the atoms are placed on a circle. Our result paves the way to implement the subwavelength transportation of light in integrated optical circuits with cold atoms.
\end{abstract}
\maketitle

There is much recent development of integrated optical circuits localising and interfacing cold atoms in which quantum networks, quantum many-body physics, and quantum optics can be explored \cite{Bajcsy:2011de,Bajcsy:2009tj,Hung:2013by,Goban:2014eq}. 
This new interdisciplinary toolbox ranging from atomic physics and quantum optics to nanophotonics opens the new door to investigate a wide range of remarkable phenomena such as the coherent transportation of photons emitted from an atom \cite{Dzsotjan:2010bj,Fan:2005jk} and atomic mirrors \cite{Chang:2012co}. 
In this Letter, we examine a waveguide made with an array of atoms along linear and circular geometries.
Nanoscale waveguides that confine, guide and manipulate the electromagnetic energy on  subwavelength scales have been actively studied as optoelectronic circuits~\cite{Ozbay:2006dz,Barnes:2003ts,Zia:2006ia}. 
The plasmon-based waveguides
such as those made with nanowires or metallic particles ~\cite{Maier:2003up, Bozhevolnyi:2001gh,Berini:2000ic,Krenn:2004jr,Tanaka:2003bv,Maier:2007ui} have very excellent lateral confinement but the propagation distance is considerably limited. This originates from the intrinsic loss of the metal which hampers the implementation of nanoscale metal plasmon waveguides. This loss is reduced when dielectric scatterers are used ~\cite{du2009:pra} via so-called $``$geometric resonances". For scatterers with  size $R$, however, the geometric resonance strictly requires the lattice spacing $a$ satisfying $2R<a<\lambda/2$ where $\lambda$ is the wavelength of the guided light. Thus, the finite size $R$ of the scatterer considerably limits the flexibility of the waveguide made by scatterers.  
It has not been possible to demonstrate waveguides with arbitrary curvature from rigorous first principles numerical calculations. Nor is the waveguiding properties of simple arrays of scatterers robust under positional fluctuations.

Here, we consider a waveguide made with an array of atoms where there is no size restriction. Instead of the $``$geometric resonance" with dielectric scatterers, we employ intrinsic $``$atomic resonance" to support a propagating mode in flexible sub-wavelength geometries in this Letter. With appropriate lattice spacing $a\sim$~0.6$\lambda_0/2\pi$ of the atomic chain where $\lambda_0$ denotes the wavelength of nearly resonant light, the \textit{external} electric field exciting the first few atoms can propagate along the finite size of the chain with minimal radiation loss. We find that the transportation is robust with respect to position fluctuation of atoms and is possible when the atoms are placed on a circle.  Recent development of plasmonic nanosphere \cite{Gullans:2012kn} or two-dimensional photonic crystals~\cite{2015NaPho...9..320G} localising atoms should allow one to prepare an array of atoms with tunable lattice spacing. 

Through the excitation in an atom at one end of the chain, energy is transfered to all adjacent atoms in turn through the resonant excitations. For the chain of atoms, resonant excitations in different atoms are coupled by electromagnetic interaction and form collective optical modes. We consider the transportation of this excitation along the atomic chain is through an intrinsic p-wave resonance. Then, we investigate how excitation of  the first atom by an external electric field {\bf E}  can travel down the chain.

We expand the field near each atom locally in terms of vector spherical harmonics and examine the scattering phase shift of electric fields from each atom as is described in Ref.~\cite{Anonymous:h-EJ17DA}. To identify the waveguide mode along the chain of atoms, we employ the multiple scattering theory describing electromagnetic interactions between the different atoms~\cite{Waterman:1965db}. Here, the scattered electric field from a single atom is related to the sum of the external electric field and the scattered fields from all the other atoms. As we see below, the waveguide mode is indicated by the undiminished propagation of the electromagnetic wave down the chain after the first few atoms are excited.

\begin{figure}[tbp]
\resizebox{5.7cm}{!}{\includegraphics{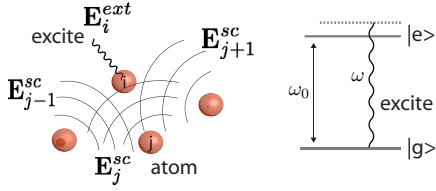}}
\caption{ (color online) Our modelled system. An array of two-level atoms responding to the external electromagnetic field $E^{ext}_i$ at the frequency of $\omega$. The $i^{th}$ atom is drived by both external fields and scattered fields $E^{SC}_j$ from $j^{th}$ atoms where $j \ne i$.}
\label{schematic_new}
\end{figure}

In the multiple scattering approach~\cite{Waterman:1965db}, the total incoming electric field driving the $i^{th}$ atom ${\bf E}^{(1)}_i$ is the sum of the external electric field ${\bf E}_i^{ext}$ and the scattered fields from all the other atoms: 
$\sum _{j\ne i}{\bf E}_{ij}^{(sc)}$, ie ${\bf E}^{(1)}_i={\bf E}_i^{ext}+ \sum _{j\ne i}{\bf E}_{ij}^{(sc)}$ as is illustrated in Fig. \ref{schematic_new}. This incoming field is scattered by the atom  and produces a scattered field ${\bf E}_{i}^{(3)}$ that is equal to the t matrix times the incoming field: ${\bf E}_{i}^{(3)}={\bf t E}_{i}^{(1)}$. The field ${\bf E}_{ij}^{(sc)}$is equal to the photon propagator ${\bf G}_{ij}$ times the scattered field from site j: ${\bf E}_{ij}^{(sc)}={\bf G_{ij}{\bf E}_{j}^{(3)}} $. Here, the t matrix is diagonal due to the rotational invariance of the Hamiltonian~\cite{Anonymous:h-EJ17DA}, and given by $t=\tan \delta/(\tan \delta+i)$ where the scattering  phase shift satisfies $\tan\delta=1/(\omega-\omega_0+i\Gamma)$ for the natural linewidth $\Gamma$ of the optical transition. To summarise, we get the set of linear equations

\begin{equation}
{\bf  E}^{(3)}_{i}= {\bf t}_i[\sum _{j\ne i}{\bf G_{ij}{\bf E}^{(3)}_{j}}+{\bf E}^{ext}_i]
\end{equation}
In momentum space, the multiple scattering equation
\begin{equation}
{\bf  E}^{(3)} (k)= \frac{{\bf E}^{ext}(k)}{1-{\bf t(k)}{\bf G(k)}}
\end{equation}
gives the resonance condition $\bf t(k)G(k)=1$.

\paragraph*{Linear Chain}To illustrate the subwavelength transport of photons, we first consider a one-dimensional array of 30-320 atoms with a p-wave resonance at an angular frequency $\omega_0$ and separated by lattice spacings $a$. The distribution of the electric field as a function of the position of the atoms is numerically calculated when the first few atoms are excited by the external electric field. We find that there can be a large response at the end of the chain but only when the external field is perpendicular to the line of atoms. We attribute this to the nature of p-wave resonance of atoms. In addition, we investigate the dependence on the way the system is excited. The propagated waveguide mode at the end of the chain does not show significant difference for two cases - when the first or the first two atoms are excited. For the rest of the part, we consider the case when the only first atom is externally excited.

\begin{figure}[tbp]
\resizebox{8.5cm}{!}{\includegraphics{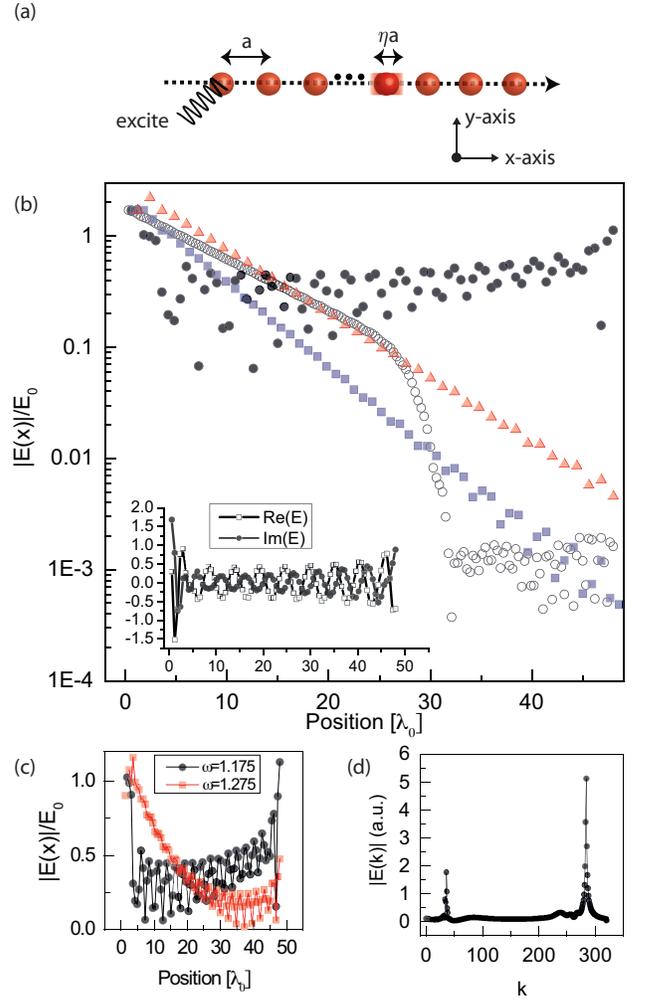}}
\caption{ (color online) Waveguide mode along the linear chain of atoms. (a) An array of atoms is placed on a straight line with the separation of $a$. Here, the amount of position fluctuations is indicated by $\eta a$. (b) The normalized intensity of the electric field $|E|$ is shown as a function of the position of atoms for different lattice spacing $a=0.3 \lambda_0/2\pi$ (open circle), $a=0.6 \lambda_0/2\pi$ (solid circle), $a=0.9 \lambda_0/2\pi$ (solid triangle), and $a=1.2 \lambda_0/2\pi$ (solid square). Here, $E_0$ is the the amplitude of the external electric field exciting the first few atoms. The inset shows the real and imaginary part of electric field for the propagating mode when $a=0.6\lambda_0/2\pi$. (c) Distribution of the electric field as a function of the position is shown for different incoming angular frequency normalized by $E_0$. (d) Fourier transform of the  electric field along the progating direction (x) as a function of the wave vector k in units of $2\pi /(aN)$ for N=320 atoms on a straight line of lattice spacing $a=0.6\lambda_0/2\pi$.  }
\label{fig2}
\end{figure}

Our numerical calculation suggests that the waveguide mode can be achieved for a spacing $a \sim 0.6 \lambda_0/2\pi$. Since the resonance is intrinsic in the atomic system , the wavelength of resonant light $\lambda_0$ is species-specific. In Fig. \ref{fig2} we show the electric field magnitude, normalized by external electric field amplitude $E_0$,  as a function of the position  in the unit of $\lambda_0$ for different lattice spacing $a$. Only for $a$ around $0.6\lambda_0/2\pi$ is there a ``waveguide" mode. For the other array spacing, the wave is non-propagating down the chain.  The real part and the imaginary part of the field as a function of atom position are compared in the inset of Fig. \ref{fig2} (b). As expected, they are of comparable magnitude and exhibit the same spatial dependence. This figure illustrates the existence of a "carrier wave" of high periodicity, which is usually modulated for different situations.

{\it Waveguide condition}
Let $k_\perp$ be the transverse wave vector normal to the propagation direction. To
suppress the radiation loss,
$k_\perp$ must be imaginary.
The transverse wave vector $k_{\perp }$ is given by $%
k_{\perp }=\sqrt{(2\pi /\lambda )^{2}-k_{x}^{2}}$, with $\lambda $ the
wavelength of the guided wave and $k_{x}$ the wave vector along propagation direction.
The  Fourier transform of the  electric field in the transverse mode is shown as a function of the wave vector k in units of $2\pi/(N a)$ for N=320 atoms (see Fig.\ref{fig2} (d)). The incoming wave has a wave vector $k_i=\omega Na/(2\pi c)=35.9$ in the present unit. This is much less than the dominant peak at $k=284$ and therefore $k_y$ is imaginary. This large periodicity is mainly due to the ``carrier wave", which sustains the subwavelength transport. There is another peak in the Fourier transform at $k_i$, which reflects the contribution from atoms at the beginning of the chain. 

As $k_{x}$ takes its maximum at the Brillouin zone boundary $K_{B}=\pi
/a$ for a chain of particles, 
the necessary condition to support the waveguide modes is~\cite{Burin:2004js,Blaustein:2007ig}
\begin{equation}
a<\lambda /2.
\end{equation}%

Nanoparticles are required to be packed as closely as possible 
regardless of waveguides made of metallic or dielectric particles \cite{Barnes:2003ts,Maier:2007ui,Quinten:1998ho}. For atoms, however, the resonance is intrinsic, there is much greater flexibility in the choice of the lattice constant.

The above discussion shows that the one-dimensional lattice spacing cannot be too big. Our numerical result indicates that the lattice spacing cannot be arbitrarily small as well. Physically, as the lattice spacing becomes small, it becomes possible to describe the system with an effective dielectric constant $\epsilon_{eff}$ that  becomes bigger and bigger as the lattice constant is decreased. Since $k_y^2=\omega^2\epsilon_{eff} /c^2-k_x^2$, it becomes more and more difficult to render $k_y^2$ negative to localize the electromagnetic wave. Mathematically we find that the imaginary part of the resonance frequency increases when the lattice spacing becomes very small. It becomes impossible to reduce the radiative loss. The details of this is described in Ref.~\cite{Anonymous:h-EJ17DA}.

\begin{figure}[tbp]
\begin{center}
\includegraphics[width=8.7cm]{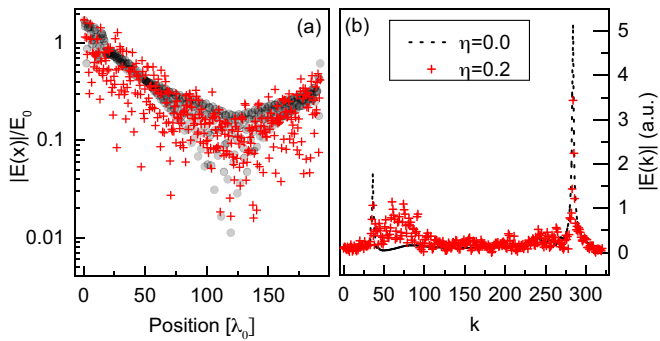}
\caption{ (color online) (a) The electric field amplitudes are shown as a function of the position for N=320 atoms of different amount of positional fluctuations: $\eta=0$ (black solid circle)and  $\eta=0.2$ (red cross). In (b), the Fourier transform of transverse electic field is shown in units of $2\pi a/N$ for N=320 atoms.}
\label{fig3}
\end{center}
\end{figure}

\paragraph*{Frequency and Length Dependence}
Now we examine the electric field intensity along the chain for different angular frequencies normalized by the resonance angular frequency of the atom as shown in Fig. \ref{fig2} (c). The maximum amplitude occurs around $\omega/\omega_0=1.175$. As the frequency moves away, the mode becomes nonpropagating, as is illustrated. We also investigated the transport through atoms for straight lines of two different lengths but the same lattice spacing. The ``carrier wave" periodicity remains the same, but the response is bigger for the longer chain. Generally, the resonance energy levels form a discrete set for a finite chain of atoms. The spacing between the levels becomes smaller as the length is increased. It is easier to find a resonance energy level closer to the driving frequency.

\paragraph*{Position fluctuation}
We next analyze the possibility of transport through atoms with fluctuations in their positions so that the deviation of the atom from its perfect position is equal to $ \eta a\xi$. Here $\xi$ is a random number between -1 and 1 and  thus $\eta$ measures the amount of the fluctuation. The dependence of the electric field $E$ on distance and its Fourier transform is shown in Fig. \ref{fig3} for different amount of randomness. The response is quite robust. The waveguide mode persists for all the cases we tested. We think this is made possible because the spacing of the lattice is much less than the wavelength and that the resonance is intrinsic and is not from a Bragg reflection.
  
To achieve the waveguide mode along the atomic chain, the controlled patterning of trapped atoms is critical. It can be done either by locating nanoparticles lithographically~\cite{Nagpal:2009jr} or through the self-assembly mechanism~\cite{Fan:2010jn}. In both methods, atoms can be confined within the subwavelength trap associated with the near-field confinement.  Another promising way is to confine atoms in a hollow-core fiber~\cite{Bajcsy:2011de,Christensen:2008tk} or in a two-dimensional photonic crystal~\cite{2015NaPho...9..320G,Hung:2013by} producing subwavelength lattice potentials.

\begin{figure}[tbp]
\includegraphics[width=8.5cm]{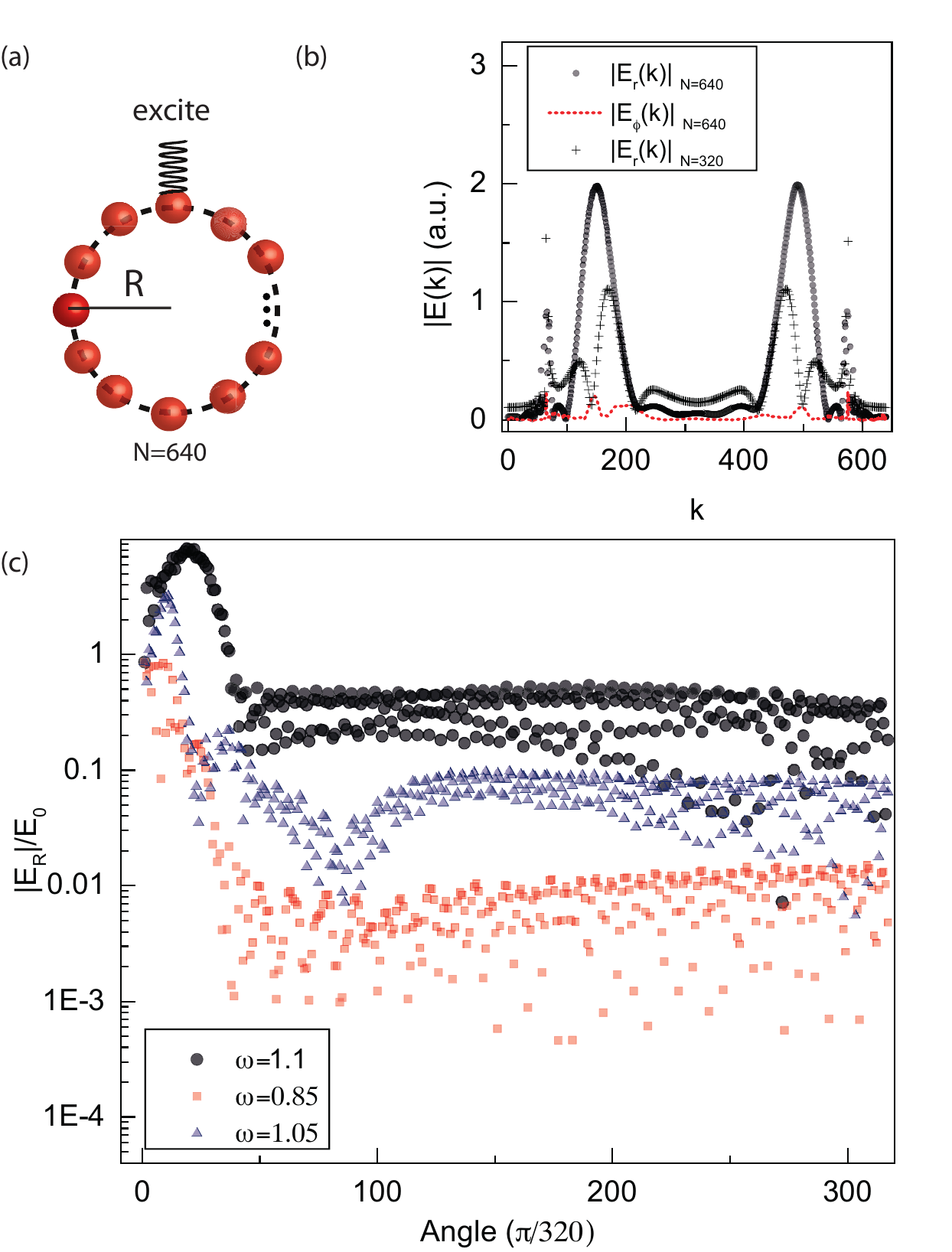}
\caption{ (color online) Waveguide mode on the circular atomic chain. (a) The atom at the top is excited by an external electric field resulting propagating modes in both directions. (b) Fourier transform of radial ($|E_r(k)|$) and tangential ($|E_{\phi}(k)|$) electric fields as a function of the wave vector k in units of $2\pi a/N$ for N=640 atoms on the circumferences of circle at an angular frequency $\omega=1.1$.
(c) Distribution of the radial electric field as a function of the position of the atoms normalized by the lattice constant $a = 0.6 \lambda_0/2\pi$ for atoms on the circumferences of a circle with 640 atoms at different frequencies.}
\label{fig4}
\end{figure}

In our model, we consider an array of atoms retaining a single two-level optical transition for supporting a waveguide mode. Among alkali or alkali-earth-like atoms, for example, ytterbium or strontium isotopes have a single principle optical transition due to the total electron spin of zero~\cite{Sansonetti:2005ul}. For ytterbium atoms, an ideal separation between atoms found above is  $0.6\lambda_{Yb}/2\pi \sim$40nm (55nm) where the $^1S_0-{^1}P_1$ ($^1S_0-{^3}P_1$) optical transition has the wavelength $\lambda_{Yb}$ of 399~nm (556~nm).  Our analysis may be extended to the array of atoms with more than one principle resonance such as rubidium, sodium and potassium. 

\paragraph*{Circular chain}
Finally we consider ``cavities" made from atoms arranged on a circle. For geometric resonances, we have not been able to construct circular cavities with small radiative loss. For intrinsic resonances with smaller lattice spacings, this has now become possible.
The distribution of the electric field intensity in real space along the circle is shown in 
Fig. \ref{fig4} (c) for different frequencies  with the external field along different directions. The resonance frequency for propagation is now shifted from that for the straight line, the dependence on the resonance frequency is quite sharp. The mode is excited no matter what direction the external field is.

 In Fig. \ref{fig4} (b) we show the Fourier transforms of the electric fields in the radial and the tangential directions for N=640 atoms separated by a lattice spacing $a=0.6 \lambda_0/2\pi$ on the circumferences of circle after the atom at the top is excited by an external field at an angular frequency $\omega=1.1$. As can be seen , the field is mainly along the radial direction. We have also studied the system for various numbers of atoms.  We show in Fig.~\ref{fig4} (b) the Fourier transforms of the electric fields in the radial direction for a smaller number of N=320 atoms separated by the same lattice spacing $a=0.6$ on the circumferences of circle after the atom at the top is excited by an external field. 
The magnitude of the field is now less as is expected.

We emphasize that even though this looks similar to the Whispering gallery mode~\cite{Rayleigh:1945vy}, there is no outside wall to confine the wave in the present case. Indeed, there is a high wave-vector ``carrier wave" which makes the cavity possible. In addition, the transport of photons with reduced radiative loss has been  investigated also in different systems. For example, a system of two distant atoms connected by a one-dimensional line waveguide \cite{Fan:2005jk} or subwavelength-diameter nanofibers~\cite{2003Natur.426..816T}  reveal interesting transport properties with reduced radiation.

In conclusion, we demonstrate that the electromagnetic energy can be efficiently transported along a chain of atoms through an atomic resonance with a wavelength of $\lambda_0$. We numerically analyze the waveguide mode propagating along a finite chain with appropriate lattice spacing $ a\sim$~0.6$\lambda_0/2\pi$. The efficient subwavelength transportation of electromagnetic energy along the one-dimensional waveguide would provide a main element in integrated nanoscale optical circuits. We believe that ultracold atoms can be localised with subwavelength lattice spacing in the plasmonic nanosphere traps or hollow fibers. Our study of 1D array structure represents a basic element of complex geometries such as two-dimensional patterning of atoms. For example,  it should be interesting to study how the light propagates in the two-dimensional triangular array from one end to the other~\cite{Becker:2010de}. 

The work was supported by the Hong Kong Research Grants Council (Project No. ECS26300014). STC thanks the hospitability of the Institute for advanced studies  and the physics department of Hong Kong University of Science and Technology where this work was carried out.

\clearpage
\newpage

\begin{center}
\Large \textbf{Supplementary Information}
\end{center}

\section*{1. Coupling to atoms}

We have the usual Hamiltonian $H=-e{\bf r\cdot E}.$ 
The electric field can be expanded in terms of vector spherical harmonics:
\begin{eqnarray}
\mathbf{E^J(r)} &=&\sum_{l=0}\sum_{m=-l}^{l}a_{lm}^{JE}{N}%
_{lm}^{(J)}+a_{lm}^{JH}{M}_{lm}^{(J)}  
\end{eqnarray}
The  vector spherical harmonics are defined as%
\begin{eqnarray}
\mathbf{M}_{mn}^{(J)}(x,\theta ,\phi ) &=&\frac{\gamma _{mn}}{k}\nabla
\times \lbrack \mathbf{r}\psi _{mn}^{(J)}(x,\theta ,\phi )] \\
\mathbf{N}_{mn}^{(J)}(x,\theta ,\phi ) &=&\frac{\gamma _{mn}}{k}\nabla
\times \mathbf{M}_{mn}^{(J)}(x,\theta ,\phi ) \\
\end{eqnarray}%
The scalar spherical functions are given by 
\begin{equation}
\psi _{mn}^{(J)}(x,\theta ,\phi )=z_{n}^{(J)}(x)P_{n}^{m}(\cos \theta
)e^{im\phi }
\end{equation}%
where 
\begin{equation}
z_{n}^{(J)}(x)=\left\{ 
\begin{array}{ll}
j_{n}(x) & J=1 \\ 
h_{n}^{(1)}(x) & J=3%
\end{array}%
\right.
\end{equation}%
with $x=kr$ and $n=0,1,2,...,m=0,\pm 1,...,$ $\pm n$. The superscripts J=1 and J=3 correspond to incoming and 
outgoing waves.
$x_{0}=kr_{0}$, $\gamma _{mn}=\sqrt{\frac{2n+1}{4\pi n(n+1)}\frac{%
(n-m)!}{(n+m)!}}$, 
The electromagnetic field locally is a linear combination of ${\bf M} \propto {\bf L} Y_{lm}$ and ${\bf N}_{lm} \propto \nabla\times {\bf M}_{lm}$
These  functions all have {\bf total} (orbital plus spin) angular momentum $l$.

The Hamiltonian is $H=-e{\bf r}\cdot {\bf E}.$
We have ${\bf r}\cdot {\bf \nabla} Y_{lm}=0,$ ${\bf r}\cdot {\bf L}Y_{lm}=0,$
${\bf r}\cdot {\bf r}Y_{lm}=r^2 Y_{lm}.$ Thus  ${\bf r\cdot N}_{1m}$ is the only nonzero term in the dipole approximation. ${\bf M}\propto {\bf L}Y_{lm}$ contains no radial component, thus this term is not coupled to the atom. 
To illustrate the physics we consider atoms that exhibit a p wave resonance so that 
the atom goes from a state with total angular momentum 0, $|0>$ to a state with angular momentum 1, $|1,m>$ and then reemits a photon. The tangent of the corresponding scattering phase shift  exhibit a frequency dependence given by $\tan\delta=1/(\omega-\omega_0+i\Gamma).$ The expansion coefficients of the incoming and outgoing fields are related by the {\bf t} matrix : ${\bf a}^{(3)}={\bf t}{\bf a}^{(1)}.$ Because the Hamiltonian only involves the electric field and not the magnetic field, there  is no coupling involving the coefficients ${\bf a}^{(JH)}.$ Since we assume a ground state of zero angular momentum and since the Hamiltonian is rotationally invariant, the t matrix is diagonal in $l,m$. The t matrix is related to the tangent of the scattering phase shift $\delta$ by $t=\tan \delta/(\tan \delta+i).$
For this resonance
the coupling to the electric field is proportional to $<0|{\bf E }\cdot{\bf r}|1,m>.$ Since ${\bf N}_{lm}$ is of total angular momentum l, by the Wigner Eckert theorem, only the terms with l=1 comes in.
 
\section*{2. Multiple scattering}
The response of the lines  of atoms to an external electromagnetic field is treated by multiple scattering theory. The total incoming electric field driving the $i^{th}$ atom ${\bf E}^{(1)}_i$ is the sum of the external electric field ${\bf E}_i^{ext}$ and the scattered fields from all the other atoms: 
$\sum _{j\ne i}{\bf E}_{ij}^{(sc)}$, ie ${\bf E}^{(1)}_i={\bf E}_i^{ext}+ \sum _{j\ne i}{\bf E}_{ij}^{(sc)}.$ This incoming field is scattered by the atom  and produces a scattered field ${\bf E}_{i}^{(3)}$ that is equal to the t matrix times the incoming field: ${\bf E}_{i}^{(3)}={\bf t E}_{i}^{(1)}$.  The field ${\bf E}_{ij}^{(sc)}$is equal to the photon propagator ${\bf G}_{ij}$ times the scattered field from site j: ${\bf E}_{ij}^{(sc)}={\bf G_{ij}{\bf E}_{j}^{(3)}}. $ To summarise, we get the set of linear equations
\begin{equation}
{\bf  E}_{i}^{(3)}- \sum _{j\ne i}{\bf G_{ij}{\bf t E}_{j}^{(3)}}={\bf t E}_i^{ext}
\end{equation}


The photon propagator~\cite{Mandl:2013wb} in terms of the vector potential $A_{\mu}$ is $D_{\mu\nu}=-g_{\mu\nu}/(k^2+i\epsilon).$
Because $\phi=0,$ we need not worry about $\mu=\nu=0.$ 
In classical E\&M, we worry about the Green's function $G(r)=\int d^3r
/(k^2-\omega^2).$ This is identical to the Fourier transform of the 
photon propagator.

\section*{3. Small lattice spacing}
In this appendix, we describe the physics when the lattice spacing is small.
The multiple scattering equation is:
\begin{equation}
{\bf  E}_{i}^{(3)}= {\bf t}_i[\sum _{j\ne i}{\bf G_{ij}{\bf E}_{j}^{(3)}}+{\bf E}_i^{ext}]
\end{equation}

We consider a line of atoms line up along the z axis. Then the Green's function is particularly simple\cite{Xu:1995ef}.
$G$ is diagonal in m and is given by
$$G_{ij}^{n,\nu}=C_0\sum_{q=0}^{min(n,\nu)} i^pC'_pa_qh_p(k_0|R_{ij}|)$$ for $R$ on both the left and the right.
where $p=n+\nu-2q,$ $C'_p=n(n+1)+\nu(\nu+1)-p(p+1),$ $a$ is the Wigner 3j symbol. For the present approximation, $n=\nu=1$. We thus get ($C_2'=4-6=-2,$ $C_0'=4.$
$$G_{ij}^{1,1}=C_0[ 2a_0h_2(k_0R_{ij})+4a_1h_0(k_0R_{ij})]$$,
For an infinite chain, $E_l\propto \exp(ikla).$
We need $$G(k)=\sum_{R\neq 0}G(R)=C_0[ 2a_0g_2(k)+4a_1g_0(k)]$$
where $g_n(k)=\sum_{R\neq 0} \exp(ikR)h_n(k_0R)$
The above equation becomes 
\begin{equation}
E(k)= t[ G(k) E(k)+E^{ext}(k)]
\end{equation}
The resonance condition is 
$$1=tG$$

We look at the small a limit. Take $h=j+iy.$ $h_0(x)=(\sin(x)-i\cos(x))/x=-i[i\sin(x)+\cos(x))/x=-i\exp(ix)/x$
Define $x_j=ja$. For example, for n=0, $h_0(x)=-i\exp(ix)/x,$
$$g_0(k)=2\sum_{j\neq 0} (\Delta x_j)\exp(ikx_j)h_0(k_0|x_j|)/a$$
$$=\sum_{j> 0} (\Delta x_j)(\exp[i(k+k_0)x_j]+\exp[i(-k+k_0)x_j])/(k_0x_ja)$$
$$\approx \int_a^{\infty} dx (\exp[i(k+k_0)x]+\exp[i(-k+k_0)x])/(k_0xa).$$
Both the real and the imaginary part of $g$ scales as $ln(a)/a$. This dependence will also dominate $1-tG$.
Thus the imaginary part of $E_s=tE_i/(1-tG)$ will not be small in the small a limit.


\begin{thebibliography}{33}
\expandafter\ifx\csname natexlab\endcsname\relax\def\natexlab#1{#1}\fi
\expandafter\ifx\csname bibnamefont\endcsname\relax
  \def\bibnamefont#1{#1}\fi
\expandafter\ifx\csname bibfnamefont\endcsname\relax
  \def\bibfnamefont#1{#1}\fi
\expandafter\ifx\csname citenamefont\endcsname\relax
  \def\citenamefont#1{#1}\fi
\expandafter\ifx\csname url\endcsname\relax
  \def\url#1{\texttt{#1}}\fi
\expandafter\ifx\csname urlprefix\endcsname\relax\def\urlprefix{URL }\fi
\providecommand{\bibinfo}[2]{#2}
\providecommand{\eprint}[2][]{\url{#2}}

\bibitem[{\citenamefont{Bajcsy et~al.}(2011)\citenamefont{Bajcsy, Hofferberth,
  Peyronel, Balic, Liang, Zibrov, Vuletic, and Lukin}}]{Bajcsy:2011de}
\bibinfo{author}{\bibfnamefont{M.}~\bibnamefont{Bajcsy}},
  \bibinfo{author}{\bibfnamefont{S.}~\bibnamefont{Hofferberth}},
  \bibinfo{author}{\bibfnamefont{T.}~\bibnamefont{Peyronel}},
  \bibinfo{author}{\bibfnamefont{V.}~\bibnamefont{Balic}},
  \bibinfo{author}{\bibfnamefont{Q.}~\bibnamefont{Liang}},
  \bibinfo{author}{\bibfnamefont{A.~S.} \bibnamefont{Zibrov}},
  \bibinfo{author}{\bibfnamefont{V.}~\bibnamefont{Vuletic}}, \bibnamefont{and}
  \bibinfo{author}{\bibfnamefont{M.~D.} \bibnamefont{Lukin}},
  \bibinfo{journal}{Physical Review A} \textbf{\bibinfo{volume}{83}},
  \bibinfo{pages}{063830} (\bibinfo{year}{2011}).

\bibitem[{\citenamefont{Bajcsy et~al.}(2009)\citenamefont{Bajcsy, Hofferberth,
  Balic, Peyronel, Hafezi, Zibrov, Vuletic, and Lukin}}]{Bajcsy:2009tj}
\bibinfo{author}{\bibfnamefont{M.}~\bibnamefont{Bajcsy}},
  \bibinfo{author}{\bibfnamefont{S.}~\bibnamefont{Hofferberth}},
  \bibinfo{author}{\bibfnamefont{V.}~\bibnamefont{Balic}},
  \bibinfo{author}{\bibfnamefont{T.}~\bibnamefont{Peyronel}},
  \bibinfo{author}{\bibfnamefont{M.}~\bibnamefont{Hafezi}},
  \bibinfo{author}{\bibfnamefont{A.~S.} \bibnamefont{Zibrov}},
  \bibinfo{author}{\bibfnamefont{V.}~\bibnamefont{Vuletic}}, \bibnamefont{and}
  \bibinfo{author}{\bibfnamefont{M.~D.} \bibnamefont{Lukin}},
  \bibinfo{journal}{Physical Review Letters} \textbf{\bibinfo{volume}{102}},
  \bibinfo{pages}{203902} (\bibinfo{year}{2009}).

\bibitem[{\citenamefont{Hung et~al.}(2013)\citenamefont{Hung, Meenehan, Chang,
  Painter, and Kimble}}]{Hung:2013by}
\bibinfo{author}{\bibfnamefont{C.~L.} \bibnamefont{Hung}},
  \bibinfo{author}{\bibfnamefont{S.~M.} \bibnamefont{Meenehan}},
  \bibinfo{author}{\bibfnamefont{D.~E.} \bibnamefont{Chang}},
  \bibinfo{author}{\bibfnamefont{O.}~\bibnamefont{Painter}}, \bibnamefont{and}
  \bibinfo{author}{\bibfnamefont{H.~J.} \bibnamefont{Kimble}},
  \bibinfo{journal}{New Journal of Physics} \textbf{\bibinfo{volume}{15}},
  \bibinfo{pages}{083026} (\bibinfo{year}{2013}).

\bibitem[{\citenamefont{Goban et~al.}(2014)\citenamefont{Goban, Hung, Yu, Hood,
  Muniz, Lee, Martin, McClung, Choi, Chang et~al.}}]{Goban:2014eq}
\bibinfo{author}{\bibfnamefont{A.}~\bibnamefont{Goban}},
  \bibinfo{author}{\bibfnamefont{C.~L.} \bibnamefont{Hung}},
  \bibinfo{author}{\bibfnamefont{S.~P.} \bibnamefont{Yu}},
  \bibinfo{author}{\bibfnamefont{J.~D.} \bibnamefont{Hood}},
  \bibinfo{author}{\bibfnamefont{J.~A.} \bibnamefont{Muniz}},
  \bibinfo{author}{\bibfnamefont{J.~H.} \bibnamefont{Lee}},
  \bibinfo{author}{\bibfnamefont{M.~J.} \bibnamefont{Martin}},
  \bibinfo{author}{\bibfnamefont{A.~C.} \bibnamefont{McClung}},
  \bibinfo{author}{\bibfnamefont{K.~S.} \bibnamefont{Choi}},
  \bibinfo{author}{\bibfnamefont{D.~E.} \bibnamefont{Chang}},
  \bibnamefont{et~al.}, \bibinfo{journal}{Nature Communications}
  \textbf{\bibinfo{volume}{5}} (\bibinfo{year}{2014}).

\bibitem[{\citenamefont{Dzsotjan et~al.}(2010)\citenamefont{Dzsotjan,
  S{\o}rensen, and Fleischhauer}}]{Dzsotjan:2010bj}
\bibinfo{author}{\bibfnamefont{D.}~\bibnamefont{Dzsotjan}},
  \bibinfo{author}{\bibfnamefont{A.~S.} \bibnamefont{S{\o}rensen}},
  \bibnamefont{and}
  \bibinfo{author}{\bibfnamefont{M.}~\bibnamefont{Fleischhauer}},
  \bibinfo{journal}{Physical Review B} \textbf{\bibinfo{volume}{82}},
  \bibinfo{pages}{075427} (\bibinfo{year}{2010}).

\bibitem[{\citenamefont{Fan}(2005)}]{Fan:2005jk}
\bibinfo{author}{\bibfnamefont{J.~T. S.~S.} \bibnamefont{Fan}},
  \bibinfo{journal}{Optics Letters} \textbf{\bibinfo{volume}{30}},
  \bibinfo{pages}{2001} (\bibinfo{year}{2005}).

\bibitem[{\citenamefont{Chang et~al.}(2012)\citenamefont{Chang, Jiang,
  Gorshkov, and Kimble}}]{Chang:2012co}
\bibinfo{author}{\bibfnamefont{D.~E.} \bibnamefont{Chang}},
  \bibinfo{author}{\bibfnamefont{L.}~\bibnamefont{Jiang}},
  \bibinfo{author}{\bibfnamefont{A.~V.} \bibnamefont{Gorshkov}},
  \bibnamefont{and} \bibinfo{author}{\bibfnamefont{H.~J.}
  \bibnamefont{Kimble}}, \bibinfo{journal}{New Journal of Physics}
  \textbf{\bibinfo{volume}{14}}, \bibinfo{pages}{063003}
  (\bibinfo{year}{2012}).

\bibitem[{\citenamefont{Ozbay}(2006)}]{Ozbay:2006dz}
\bibinfo{author}{\bibfnamefont{E.}~\bibnamefont{Ozbay}},
  \bibinfo{journal}{Science} \textbf{\bibinfo{volume}{311}},
  \bibinfo{pages}{189} (\bibinfo{year}{2006}).

\bibitem[{\citenamefont{Barnes et~al.}(2003)\citenamefont{Barnes, Dereux, and
  Ebbesen}}]{Barnes:2003ts}
\bibinfo{author}{\bibfnamefont{W.~L.} \bibnamefont{Barnes}},
  \bibinfo{author}{\bibfnamefont{A.}~\bibnamefont{Dereux}}, \bibnamefont{and}
  \bibinfo{author}{\bibfnamefont{T.~W.} \bibnamefont{Ebbesen}},
  \bibinfo{journal}{Nature} \textbf{\bibinfo{volume}{424}},
  \bibinfo{pages}{824} (\bibinfo{year}{2003}).

\bibitem[{\citenamefont{Zia et~al.}(2006)\citenamefont{Zia, Schuller, Chandran,
  and Brongersma}}]{Zia:2006ia}
\bibinfo{author}{\bibfnamefont{R.}~\bibnamefont{Zia}},
  \bibinfo{author}{\bibfnamefont{J.~A.} \bibnamefont{Schuller}},
  \bibinfo{author}{\bibfnamefont{A.}~\bibnamefont{Chandran}}, \bibnamefont{and}
  \bibinfo{author}{\bibfnamefont{M.~L.} \bibnamefont{Brongersma}},
  \bibinfo{journal}{Materials today} \textbf{\bibinfo{volume}{9}},
  \bibinfo{pages}{20} (\bibinfo{year}{2006}).

\bibitem[{\citenamefont{Maier et~al.}(2003)\citenamefont{Maier, Kik, Atwater,
  Meltzer, Harel, Koel, and Requicha}}]{Maier:2003up}
\bibinfo{author}{\bibfnamefont{S.~A.} \bibnamefont{Maier}},
  \bibinfo{author}{\bibfnamefont{P.~G.} \bibnamefont{Kik}},
  \bibinfo{author}{\bibfnamefont{H.~A.} \bibnamefont{Atwater}},
  \bibinfo{author}{\bibfnamefont{S.}~\bibnamefont{Meltzer}},
  \bibinfo{author}{\bibfnamefont{E.}~\bibnamefont{Harel}},
  \bibinfo{author}{\bibfnamefont{B.~E.} \bibnamefont{Koel}}, \bibnamefont{and}
  \bibinfo{author}{\bibfnamefont{A.~A.~G.} \bibnamefont{Requicha}},
  \bibinfo{journal}{Nature Materials} \textbf{\bibinfo{volume}{2}},
  \bibinfo{pages}{229} (\bibinfo{year}{2003}).

\bibitem[{\citenamefont{Bozhevolnyi et~al.}(2001)\citenamefont{Bozhevolnyi,
  Erland, Leosson, Skovgaard, and Hvam}}]{Bozhevolnyi:2001gh}
\bibinfo{author}{\bibfnamefont{S.~I.} \bibnamefont{Bozhevolnyi}},
  \bibinfo{author}{\bibfnamefont{J.}~\bibnamefont{Erland}},
  \bibinfo{author}{\bibfnamefont{K.}~\bibnamefont{Leosson}},
  \bibinfo{author}{\bibfnamefont{P.~M.~W.} \bibnamefont{Skovgaard}},
  \bibnamefont{and} \bibinfo{author}{\bibfnamefont{J.~M.} \bibnamefont{Hvam}},
  \bibinfo{journal}{Physical Review Letters} \textbf{\bibinfo{volume}{86}},
  \bibinfo{pages}{3008} (\bibinfo{year}{2001}).

\bibitem[{\citenamefont{Berini}(2000)}]{Berini:2000ic}
\bibinfo{author}{\bibfnamefont{P.}~\bibnamefont{Berini}},
  \bibinfo{journal}{Physical Review B} \textbf{\bibinfo{volume}{61}},
  \bibinfo{pages}{10484} (\bibinfo{year}{2000}).

\bibitem[{\citenamefont{Krenn and Weeber}(2004)}]{Krenn:2004jr}
\bibinfo{author}{\bibfnamefont{J.~R.} \bibnamefont{Krenn}} \bibnamefont{and}
  \bibinfo{author}{\bibfnamefont{J.~C.} \bibnamefont{Weeber}},
  \bibinfo{journal}{Philos.Trans.R. Soc.London, Ser.A}
  \textbf{\bibinfo{volume}{362}}, \bibinfo{pages}{739} (\bibinfo{year}{2004}).

\bibitem[{\citenamefont{Tanaka and Tanaka}(2003)}]{Tanaka:2003bv}
\bibinfo{author}{\bibfnamefont{K.}~\bibnamefont{Tanaka}} \bibnamefont{and}
  \bibinfo{author}{\bibfnamefont{M.}~\bibnamefont{Tanaka}},
  \bibinfo{journal}{Applied Physics Letters} \textbf{\bibinfo{volume}{82}},
  \bibinfo{pages}{1158} (\bibinfo{year}{2003}).

\bibitem[{\citenamefont{Maier}(2007)}]{Maier:2007ui}
\bibinfo{author}{\bibfnamefont{S.~A.} \bibnamefont{Maier}},
  \emph{\bibinfo{title}{{Plasmonics: fundamentals and applications:
  fundamentals and applications}}} (\bibinfo{year}{2007}).

\bibitem[{\citenamefont{Du et~al.}(2009)\citenamefont{Du, Liu, Lin, Zi, and
  Chui}}]{du2009:pra}
\bibinfo{author}{\bibfnamefont{J.}~\bibnamefont{Du}},
  \bibinfo{author}{\bibfnamefont{S.}~\bibnamefont{Liu}},
  \bibinfo{author}{\bibfnamefont{Z.}~\bibnamefont{Lin}},
  \bibinfo{author}{\bibfnamefont{J.}~\bibnamefont{Zi}}, \bibnamefont{and}
  \bibinfo{author}{\bibfnamefont{S.~T.} \bibnamefont{Chui}},
  \bibinfo{journal}{Physical Review A} \textbf{\bibinfo{volume}{79}},
  \bibinfo{pages}{051801} (\bibinfo{year}{2009}).

\bibitem[{\citenamefont{Gullans et~al.}(2012)\citenamefont{Gullans, Tiecke,
  Chang, and Feist}}]{Gullans:2012kn}
\bibinfo{author}{\bibfnamefont{M.}~\bibnamefont{Gullans}},
  \bibinfo{author}{\bibfnamefont{T.~G.} \bibnamefont{Tiecke}},
  \bibinfo{author}{\bibfnamefont{D.~E.} \bibnamefont{Chang}}, \bibnamefont{and}
  \bibinfo{author}{\bibfnamefont{J.}~\bibnamefont{Feist}},
  \bibinfo{journal}{Physical Review}  (\bibinfo{year}{2012}).

\bibitem[{\citenamefont{Gonz{\'a}lez-Tudela
  et~al.}(2015)\citenamefont{Gonz{\'a}lez-Tudela, Hung, Chang, Cirac, and
  Kimble}}]{2015NaPho...9..320G}
\bibinfo{author}{\bibfnamefont{A.}~\bibnamefont{Gonz{\'a}lez-Tudela}},
  \bibinfo{author}{\bibfnamefont{C.~L.} \bibnamefont{Hung}},
  \bibinfo{author}{\bibfnamefont{D.~E.} \bibnamefont{Chang}},
  \bibinfo{author}{\bibfnamefont{J.}~\bibnamefont{Cirac}}, \bibnamefont{and}
  \bibinfo{author}{\bibfnamefont{H.~J.} \bibnamefont{Kimble}},
  \bibinfo{journal}{Nature Photonics} \textbf{\bibinfo{volume}{9}},
  \bibinfo{pages}{320} (\bibinfo{year}{2015}).

\bibitem[{Ano()}]{Anonymous:h-EJ17DA}
\bibinfo{journal}{See supplemental material}  (????).

\bibitem[{\citenamefont{Waterman}(1965)}]{Waterman:1965db}
\bibinfo{author}{\bibfnamefont{P.~C.} \bibnamefont{Waterman}},
  \bibinfo{journal}{Proceedings of the IEEE} \textbf{\bibinfo{volume}{53}},
  \bibinfo{pages}{805} (\bibinfo{year}{1965}).

\bibitem[{\citenamefont{Burin et~al.}(2004)\citenamefont{Burin, Cao, Schatz,
  and Ratner}}]{Burin:2004js}
\bibinfo{author}{\bibfnamefont{A.~L.} \bibnamefont{Burin}},
  \bibinfo{author}{\bibfnamefont{H.}~\bibnamefont{Cao}},
  \bibinfo{author}{\bibfnamefont{G.~C.} \bibnamefont{Schatz}},
  \bibnamefont{and} \bibinfo{author}{\bibfnamefont{M.~A.}
  \bibnamefont{Ratner}}, \bibinfo{journal}{JOSA B}
  \textbf{\bibinfo{volume}{21}}, \bibinfo{pages}{121} (\bibinfo{year}{2004}).

\bibitem[{\citenamefont{Blaustein et~al.}(2007)\citenamefont{Blaustein, Gozman,
  Samoylova, Polishchuk, and Burin}}]{Blaustein:2007ig}
\bibinfo{author}{\bibfnamefont{G.~S.} \bibnamefont{Blaustein}},
  \bibinfo{author}{\bibfnamefont{M.~I.} \bibnamefont{Gozman}},
  \bibinfo{author}{\bibfnamefont{O.}~\bibnamefont{Samoylova}},
  \bibinfo{author}{\bibfnamefont{I.~Y.} \bibnamefont{Polishchuk}},
  \bibnamefont{and} \bibinfo{author}{\bibfnamefont{A.~L.} \bibnamefont{Burin}},
  \bibinfo{journal}{Optics Express} \textbf{\bibinfo{volume}{15}},
  \bibinfo{pages}{17380} (\bibinfo{year}{2007}).

\bibitem[{\citenamefont{Quinten et~al.}(1998)\citenamefont{Quinten, Leitner,
  Krenn, and Aussenegg}}]{Quinten:1998ho}
\bibinfo{author}{\bibfnamefont{M.}~\bibnamefont{Quinten}},
  \bibinfo{author}{\bibfnamefont{A.}~\bibnamefont{Leitner}},
  \bibinfo{author}{\bibfnamefont{J.~R.} \bibnamefont{Krenn}}, \bibnamefont{and}
  \bibinfo{author}{\bibfnamefont{F.~R.} \bibnamefont{Aussenegg}},
  \bibinfo{journal}{Optics Letters} \textbf{\bibinfo{volume}{23}},
  \bibinfo{pages}{1331} (\bibinfo{year}{1998}).

\bibitem[{\citenamefont{Nagpal et~al.}(2009)\citenamefont{Nagpal, Lindquist,
  Oh, and Norris}}]{Nagpal:2009jr}
\bibinfo{author}{\bibfnamefont{P.}~\bibnamefont{Nagpal}},
  \bibinfo{author}{\bibfnamefont{N.~C.} \bibnamefont{Lindquist}},
  \bibinfo{author}{\bibfnamefont{S.-H.} \bibnamefont{Oh}}, \bibnamefont{and}
  \bibinfo{author}{\bibfnamefont{D.~J.} \bibnamefont{Norris}},
  \bibinfo{journal}{Science} \textbf{\bibinfo{volume}{325}},
  \bibinfo{pages}{594} (\bibinfo{year}{2009}).

\bibitem[{\citenamefont{Fan et~al.}(2010)\citenamefont{Fan, Wu, Bao, Bao,
  Bardhan, Halas, Manoharan, Nordlander, Shvets, and Capasso}}]{Fan:2010jn}
\bibinfo{author}{\bibfnamefont{J.~A.} \bibnamefont{Fan}},
  \bibinfo{author}{\bibfnamefont{C.}~\bibnamefont{Wu}},
  \bibinfo{author}{\bibfnamefont{K.}~\bibnamefont{Bao}},
  \bibinfo{author}{\bibfnamefont{J.}~\bibnamefont{Bao}},
  \bibinfo{author}{\bibfnamefont{R.}~\bibnamefont{Bardhan}},
  \bibinfo{author}{\bibfnamefont{N.~J.} \bibnamefont{Halas}},
  \bibinfo{author}{\bibfnamefont{V.~N.} \bibnamefont{Manoharan}},
  \bibinfo{author}{\bibfnamefont{P.}~\bibnamefont{Nordlander}},
  \bibinfo{author}{\bibfnamefont{G.}~\bibnamefont{Shvets}}, \bibnamefont{and}
  \bibinfo{author}{\bibfnamefont{F.}~\bibnamefont{Capasso}},
  \bibinfo{journal}{Science} \textbf{\bibinfo{volume}{328}},
  \bibinfo{pages}{1135} (\bibinfo{year}{2010}).

\bibitem[{\citenamefont{Christensen et~al.}(2008)\citenamefont{Christensen,
  Will, Saba, Jo, Shin, Ketterle, and Pritchard}}]{Christensen:2008tk}
\bibinfo{author}{\bibfnamefont{C.~A.} \bibnamefont{Christensen}},
  \bibinfo{author}{\bibfnamefont{S.}~\bibnamefont{Will}},
  \bibinfo{author}{\bibfnamefont{M.}~\bibnamefont{Saba}},
  \bibinfo{author}{\bibfnamefont{G.-B.} \bibnamefont{Jo}},
  \bibinfo{author}{\bibfnamefont{Y.-i.} \bibnamefont{Shin}},
  \bibinfo{author}{\bibfnamefont{W.}~\bibnamefont{Ketterle}}, \bibnamefont{and}
  \bibinfo{author}{\bibfnamefont{D.}~\bibnamefont{Pritchard}},
  \bibinfo{journal}{Physical Review A} \textbf{\bibinfo{volume}{78}},
  \bibinfo{pages}{033429} (\bibinfo{year}{2008}).

\bibitem[{\citenamefont{Sansonetti and Martin}(2005)}]{Sansonetti:2005ul}
\bibinfo{author}{\bibfnamefont{J.~E.} \bibnamefont{Sansonetti}}
  \bibnamefont{and} \bibinfo{author}{\bibfnamefont{W.~C.}
  \bibnamefont{Martin}}, \bibinfo{journal}{Journal of Physical and Chemical
  Reference Data} \textbf{\bibinfo{volume}{34}}, \bibinfo{pages}{1559}
  (\bibinfo{year}{2005}).

\bibitem[{\citenamefont{Rayleigh}(1945)}]{Rayleigh:1945vy}
\bibinfo{author}{\bibfnamefont{L.}~\bibnamefont{Rayleigh}},
  \emph{\bibinfo{title}{{Rayleigh: The Theory of Sound, 1894 - Google
  Scholar}}} (\bibinfo{publisher}{Republished by Dover Publications},
  \bibinfo{year}{1945}).

\bibitem[{\citenamefont{Tong et~al.}(2003)\citenamefont{Tong, Gattass, Ashcom,
  He, Lou, Shen, Maxwell, and Mazur}}]{2003Natur.426..816T}
\bibinfo{author}{\bibfnamefont{L.}~\bibnamefont{Tong}},
  \bibinfo{author}{\bibfnamefont{R.~R.} \bibnamefont{Gattass}},
  \bibinfo{author}{\bibfnamefont{J.~B.} \bibnamefont{Ashcom}},
  \bibinfo{author}{\bibfnamefont{S.}~\bibnamefont{He}},
  \bibinfo{author}{\bibfnamefont{J.}~\bibnamefont{Lou}},
  \bibinfo{author}{\bibfnamefont{M.}~\bibnamefont{Shen}},
  \bibinfo{author}{\bibfnamefont{I.}~\bibnamefont{Maxwell}}, \bibnamefont{and}
  \bibinfo{author}{\bibfnamefont{E.}~\bibnamefont{Mazur}},
  \bibinfo{journal}{Nature} \textbf{\bibinfo{volume}{426}},
  \bibinfo{pages}{816} (\bibinfo{year}{2003}).

\bibitem[{\citenamefont{Becker et~al.}(2010)\citenamefont{Becker,
  Soltan-Panahi, Kronjager, D{\"o}rscher, Bongs, and
  Sengstock}}]{Becker:2010de}
\bibinfo{author}{\bibfnamefont{C.}~\bibnamefont{Becker}},
  \bibinfo{author}{\bibfnamefont{P.}~\bibnamefont{Soltan-Panahi}},
  \bibinfo{author}{\bibfnamefont{J.}~\bibnamefont{Kronjager}},
  \bibinfo{author}{\bibfnamefont{S.}~\bibnamefont{D{\"o}rscher}},
  \bibinfo{author}{\bibfnamefont{K.}~\bibnamefont{Bongs}}, \bibnamefont{and}
  \bibinfo{author}{\bibfnamefont{K.}~\bibnamefont{Sengstock}},
  \bibinfo{journal}{New Journal of Physics} \textbf{\bibinfo{volume}{12}},
  \bibinfo{pages}{065025} (\bibinfo{year}{2010}).

\bibitem[{\citenamefont{Mandl and Shaw}(2013)}]{Mandl:2013wb}
\bibinfo{author}{\bibfnamefont{F.}~\bibnamefont{Mandl}} \bibnamefont{and}
  \bibinfo{author}{\bibfnamefont{G.~P.} \bibnamefont{Shaw}},
  \emph{\bibinfo{title}{{Quantum Field Theory}}} (\bibinfo{publisher}{John
  Wiley {\&} Sons}, \bibinfo{year}{2013}).

\bibitem[{\citenamefont{Xu}(1995)}]{Xu:1995ef}
\bibinfo{author}{\bibfnamefont{Y.-l.} \bibnamefont{Xu}},
  \bibinfo{journal}{Applied Optics} \textbf{\bibinfo{volume}{34}},
  \bibinfo{pages}{4573} (\bibinfo{year}{1995}).

\end{thebibliography}

\end{document}